\documentclass[conference]{IEEEtran}
\usepackage{cite}
\usepackage{amsmath,amssymb,amsfonts}
\usepackage{algorithmic}
\usepackage{float}
\usepackage{graphicx}
\usepackage{textcomp}
\usepackage{xcolor}
\usepackage{lipsum}
\usepackage{subfig}
\usepackage{array}
\usepackage[hidelinks]{hyperref}
\usepackage{makecell}
\usepackage{multirow}
\usepackage{booktabs}
\usepackage[normalem]{ulem}

\hypersetup{
    colorlinks = true,
    urlcolor = blue,
    linkcolor = blue,
    citecolor = red
}

\graphicspath{ {./Figures/} }
\def\BibTeX{{\rm B\kern-.05em{\sc i\kern-.025em b}\kern-.08em
    T\kern-.1667em\lower.7ex\hbox{E}\kern-.125emX}}

\newcommand{\dt}{\operatorname{d}\!t}

\newcommand{\transpose}[1]{{#1}{}^\mathsf{T}}

\begin{document}

\title{Recursive KalmanNet: Deep Learning-Augmented Kalman Filtering for State Estimation with Consistent Uncertainty Quantification}

\author{

	\IEEEauthorblockN{
		Hassan Mortada,
		Cyril Falcon,
		Yanis Kahil,
		Mathéo Clavaud,
		Jean-Philippe Michel
	}

	\IEEEauthorblockA{
		\textit{Exail -- Navigation Systems and Applications} \\
		\textit{34 rue de la Croix de Fer, 78100 Saint Germain en Laye, France} \\
		Emails: \{firstname.lastname\}@exail.com
	}
}

\maketitle

\begin{abstract}
    State estimation in stochastic dynamical systems with noisy measurements is a challenge. While the Kalman filter is optimal for linear systems with independent Gaussian white noise, real-world conditions often deviate from these assumptions, prompting the rise of data-driven filtering techniques. This paper introduces Recursive KalmanNet, a Kalman-filter-informed recurrent neural network designed for accurate state estimation with consistent error covariance quantification. Our approach propagates error covariance using the recursive Joseph's formula and optimizes the Gaussian negative log-likelihood. Experiments with non-Gaussian measurement white noise demonstrate that our model outperforms both the conventional Kalman filter and an existing state-of-the-art deep learning based estimator.
\end{abstract}


\section{Introduction}
The Kalman Filter (KF)~\cite{Kalman} provides an optimal estimation of a state vector that evolves according to a linear differential equation, with measurements modeled as a linear combination of the state vector. The solution consists of two components: an optimal state estimate, and an associated error covariance that quantifies the uncertainty of the state estimates.

KF has been widely applied in areas such as inertial navigation~\cite{Schmidt} and robotics~\cite{chen11}. However, the KF's optimality is guaranteed only under specific conditions. First, both the state evolution and measurement models must be linear. For nonlinear systems that are well linearizable, the Extended Kalman Filter (EKF) and Unscented Kalman Filter (UKF) offer viable alternatives. Second, the noise affecting both the state evolution and measurements must be independent. Lastly, the noise must be Gaussian, white and with known covariance. Any deviation from these assumptions can result in suboptimal performance or degradation of the KF's accuracy.

The assumptions underlying the KF are often challenging to satisfy in real-world scenarios. In high-end inertial navigation system applications, for example, the state evolution equations are well approximated by linear models. However, the measurements from external sensors, such as position data obtained from GNSS (Global Navigation Satellite Systems), are rarely independent or Gaussian in nature. GNSS signals often exhibit temporal and spatial correlations due to persistent atmospheric conditions. Furthermore, environmental factors, such as whether the system operates in urban or rural areas, can introduce significant variations in signal quality, including amplitude fluctuations and distortion. In such settings, measurement noise generally dominates over state evolution noise. In practice, one can adapt the KF sub-optimally by modeling the measurement noise using a Gaussian with a relatively high variance to be robust against such error behavior.

In recent years, advancements in the field have introduced a new class of hybrid Kalman filters that combine the traditional KF structure with neural networks to mitigate the constraints of the classical approach. Mainly, the KalmanNet method~\cite{KalmanNet} replaced the Kalman gain analytical form by a Recurrent Neural Network (RNN) block trained in a supervised manner. Multiple methods~\cite{SplitKalmanNet, UncertaintyKalmanNet, CholeskyKalmanNet} derived from KalmanNet have been published to enhance it further. More details about these methods and their limitations are given in Section~\ref{SS:KalmanNetMethods}.

In this paper, we present the Recursive KalmanNet (RKN), which extends the KalmanNet framework to enhance the accuracy of state and error covariance estimates. The latter is learned by exploiting the generalized Joseph's formulation~\cite{Bucy05} for the covariance estimation in a closed-loop and recursive fashion. The main contributions of this work include an accurate state estimation with error covariance estimation which is representative of the state errors and the ability to estimate challenging time-varying gains. To the best of our knowledge, we are the first to show and quantify that our method can effectively estimate the state with a consistent error covariance. Moreover, we show that RKN outperforms the KF, KalmanNet and its derived methods in handling non-Gaussian measurement noise. 

The paper is organized as follows: Section~\ref{S:PrelStateEst} recalls the KF equations and presents KalmanNet and its derived methods. Section~\ref{S:ProposedMethod} details the proposed RKN. Section~\ref{S:Results} presents the numerical study. Section~\ref{S:Conclusion} concludes the paper.

\section{Preliminaries on state estimation}
\label{S:PrelStateEst}
This paper addresses the problem of state estimation in stochastic dynamical systems with noisy observations. Let $\mathbf{x}_{t} \in \mathbb{R}^{m}$ be the state vector and $\mathbf{z}_{t} \in \mathbb{R}^{n}$ the measurement vector, which are related through a linear state-space model:
\begin{subequations}
	\begin{align}
		\label{eq:transition}
		\mathbf{x}_t & = \mathbf{F}_t\mathbf{x}_{t-1} + \mathbf{v}_t, \\
		\label{eq:observation}
		\mathbf{z}_t & = \mathbf{H}_t\mathbf{x}_t + \mathbf{w}_t,
	\end{align}
\end{subequations}
where $\mathbf{F}_t \in \mathbb{R}^{m \times m}$ and $\mathbf{H}_t \in \mathbb{R}^{n \times m}$ are the state transition and observation matrices, respectively.
The process noise $\mathbf{v}_t$ and measurement noise $\mathbf{w}_t$ are assumed to be mutually independent, zero-mean white Gaussian noise with covariance matrices $\mathbf{Q}_t$ and $\mathbf{R}_t$, respectively. These terms, known as process noise and measurement noise, account for uncertainties and model imperfections in the system dynamics and observations. Note that in this paper we evaluate the proposed method under a challenging scenario where $\mathbf{Q}_t$ and $\mathbf{R}_t$ are unknown and the measurement noise $\mathbf{w}_t$ follows a heavy tailed distribution.

\subsection{Kalman filter}
The KF is a recursive linear estimator that provides an analytical solution to the problem of estimating the state vector of a system defined by equations \eqref{eq:transition}--\eqref{eq:observation}. It is based on a predictor-corrector scheme that estimates both the state vector and the covariance of the estimation error. KF is optimal in the minimum mean square error sense when the process and measurement noise are Gaussian, and remains the best linear estimator in the presence of non-Gaussian noise. Given an initial state $\widehat{\mathbf{x}}_{0 \vert 0}$ and error covariance $\mathbf{P}_{0 \vert 0}$, the equations for the prediction and correction steps are given below.

\subsubsection*{Prediction step}
The previous corrected state estimate $\widehat{\mathbf{x}}_{t-1 \vert t-1}$ and error covariance $\mathbf{P}_{t-1 \vert t-1}$ are propagated forward using the state transition equation \eqref{eq:transition} to obtain the predicted state estimate and error covariance:
\begin{subequations}
	\begin{align}
		\label{eq:state_pred}
		\widehat{\mathbf{x}}_{t \vert t-1} & = \mathbf{F}_t \widehat{\mathbf{x}}_{t-1 \vert t-1},                               \\
		\label{eq:covar_pred}
		\mathbf{P}_{t \vert t-1}           & = \mathbf{F}_t \mathbf{P}_{t-1 \vert t-1} \transpose{\mathbf{F}_t} + \mathbf{Q}_t.
	\end{align}
\end{subequations}

\subsubsection*{Correction step}
The measurement $\mathbf{z}_t$ is used in the observation equation \eqref{eq:observation} to correct the predicted state estimate. First, the innovation (or measurement pre-fit residual), which quantifies the discrepancy between the predicted state estimate and the measurement, along with its covariance, are computed as:
\begin{subequations}
	\begin{align}
		\label{eq:innov}
		\widehat{\mathbf{y}}_t & = \mathbf{z}_t - \mathbf{H}_t \widehat{\mathbf{x}}_{t \vert t-1},                \\
		\label{eq:innov_covar}
		\mathbf{S}_t           & = \mathbf{H}_t \mathbf{P}_{t \vert t-1} \transpose{\mathbf{H}_t} + \mathbf{R}_t.
	\end{align}
\end{subequations}
Then, the Kalman gain is derived to minimize the mean square corrected state error, given by $\mathbb{E}\left[\transpose{\mathbf{e}_t}\mathbf{e}_t\right]$, where $\mathbf{e}_t = \mathbf{x}_t - \widehat{\mathbf{x}}_{t \vert t}$. The Kalman gain is computed as:
\begin{equation}
	\label{eq:kalman_gain}
	\mathbf{K}_t  = \mathbf{P}_{t \vert t-1}\transpose{\mathbf{H}_t} \mathbf{S}_t^{-1}.
\end{equation}

Finally, the corrected state estimate and its error covariance are updated as:
\begin{subequations}
	\begin{align}
		\label{eq:state_corr}
		\widehat{\mathbf{x}}_{t \vert t} & = \widehat{\mathbf{x}}_{t \vert t-1} + \mathbf{K}_t \widehat{\mathbf{y}}_t, \\
		\label{eq:covar_corr}
		\mathbf{P}_{t \vert t}           & = (\mathbf{I} - \mathbf{K}_t \mathbf{H}_t) \mathbf{P}_{t \vert t-1},
	\end{align}
\end{subequations}
where $\mathbf{I}$ is the $m\times m$ identity matrix. The concise form of the error covariance update in \eqref{eq:covar_corr} arises from algebraic simplifications based on the Kalman gain expression in \eqref{eq:kalman_gain}.

We conclude with observations that prepare for the analysis of the numerical assessment results discussed in Section~\ref{S:Results}. The derivations presented here apply to general white noise, without assuming Gaussianity. The Kalman gain ensures that the corrected state error is uncorrelated with the innovation. However, full independence is only achieved for Gaussian noise, where the innovation becomes white noise. For this reason, the KF is the optimal mean squared error estimator for linear state-space models with independent Gaussian white noise and remains the best linear estimator, even when the Gaussian assumption is relaxed.

\subsection{Deep-learning informed Kalman filters}
\label{SS:KalmanNetMethods}
In recent years, hybrid approaches combining Kalman filtering with machine learning have been developed to address the limitations of the traditional KF. These methods mainly aim to learn the Kalman gain using supervised data-driven techniques. KalmanNet introduced in~\cite{KalmanNet} replaces the closed-form gain computation with a recurrent neural network (RNN), while preserving the KF’s predictor-corrector scheme. The RNN, comprising a Gated Recurrent Unit (GRU) and fully connected layers, is trained using a feature set and the mean squared error (MSE) as the loss function.
KalmanNet is designed to operate without prior knowledge of the noise covariances $\mathbf{Q}_t$ and $\mathbf{R}_t$, and has shown improved performance over the KF and its nonlinear extensions in cases of model mismatch or system nonlinearities. However, it does not estimate the crucial error covariance.
To address this,~\cite{UncertaintyKalmanNet} proposes estimating the error covariance and introduces a log-likelihood-based loss. Yet, this method requires the measurement matrix $\mathbf{H}_t$ to be full-column rank, limiting its generality. In~\cite{SplitKalmanNet}, two RNNs estimate the predicted error covariance $\mathbf{P}_{t\vert t-1}$ and the inverse innovation covariance ${\mathbf{S}_t}^{-1}$, using additional features such as the Jacobian of $\mathbf{H}_t$. However, the resulting error covariance is not guaranteed to be positive definite.
To overcome this,~\cite{CholeskyKalmanNet} suggests estimating the Cholesky factor of the covariance, ensuring positive definiteness. They also propose a loss function combining the MSE and the deviation between estimated and empirical covariances. However, this requires tuning a hyperparameter to balance the trade-off between estimation accuracy and covariance consistency. Additionally, the state transition matrix $\mathbf{F}_t$, capturing the system dynamic, is not incorporated into the learning process.

\begin{figure*}[t!]
	\begin{center}
		\includegraphics[width=0.71\textwidth]{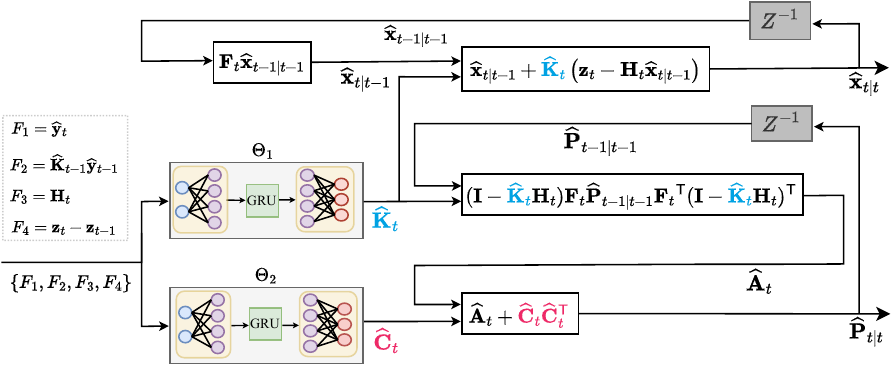}
	\end{center}
	\caption{RKN learning architecture. The gain and the corrected error covariance are estimated with two separate RNNs with set of parameters (weights and biases) $\Theta_1$ and $\Theta_2$, respectively.
		The second RNN estimates the Cholesky factor matrix of the corrected noise covariance term $\widehat{\mathbf{B}}_t$ derived from Joseph's formulation defined in~\eqref{eq:corr_noise_covar}.}
	\label{fig:architecture}
\end{figure*}

\section{Recursive KalmanNet}
\label{S:ProposedMethod}
We introduce Recursive KalmanNet (RKN), a deep learning model informed by Kalman filtering for gain estimation and consistent corrected error covariance. It operates without prior knowledge of the noise covariance matrices $\mathbf{Q}_t$ and $\mathbf{R}_t$ and does not require estimating the predicted error covariance. Instead, it updates the predicted error covariance in a single step using Joseph’s formula for the general error covariance update~\cite{Bucy05}.

RKN consists of two separate RNNs: one dedicated to gain estimation and the other to estimating the Cholesky factor of the noise-dependent term in the one-step covariance update. The overall architecture is illustrated in Fig.~\ref{fig:architecture}.

Training is performed in a supervised manner using the Gaussian negative log-likelihood of the error, ensuring uniform optimization of both the state estimates and the error covariance. Further details on the model features, learning process, and loss functions are provided below. \footnote{RKN code available on GitHub: \href{https://github.com/ixblue/RecursiveKalmanNet}{github.com/ixblue/RecursiveKalmanNet}.}

\subsection{Features}
\label{ss:features}
Gain estimation in the KF relies on the innovation covariance, which in turn depends on the noise covariances $\mathbf{Q}_t$ and $\mathbf{R}_t$. Unlike the KF, RKN does not have access to these matrices and must instead infer noise statistics from data-driven features.
RKN consists of two RNNs, both using GRUs between fully connected layers. At each time step $t$, they process the same set of input features:
\begin{itemize}
	\item $\emph{F}_1$: Innovation $\widehat{\mathbf{y}}_t$ at $t$;
	\item $\emph{F}_2$: State correction $\widehat{\mathbf{K}}_{t-1} \widehat{\mathbf{y}}_{t-1}$ from $t-1$;
	\item $\emph{F}_3$: Jacobian $\mathbf{H}_t$ of the observation equation at $t$;
	\item $\emph{F}_4$: Measurement temporal difference $\mathbf{z}_t - \mathbf{z}_{t-1}$ at $t$.
\end{itemize}
Features $\emph{F}_1$ and $\emph{F}_2$ reflect state estimation errors, while $\emph{F}_3$ and $\emph{F}_4$ provide information about measurement characteristics. Features $\emph{F}_1$, $\emph{F}_2$, and $\emph{F}_4$ are used in~\cite{KalmanNet}, and $\emph{F}_4$ appeared in~\cite{SplitKalmanNet}. 
Feature batch normalization was removed during training, as it suppresses transient dynamics in features induced by time-varying state-space models, hindering the network's ability to capture time-dependent behaviors.

Empirically, the use of squared features improves performance. This is consistent with the quadratic nature of error covariances, which the estimated parameters depend on. For instance, the innovation covariance $\mathbf{S}_t$ in the gain equation~\eqref{eq:kalman_gain} is a second-order statistic of $\emph{F}_1$.

\subsection{Error covariance with Joseph's formula}
The gain produced by the first RNN does not necessarily align with the analytical form in equation~\eqref{eq:kalman_gain}, potentially invalidating equation~\eqref{eq:covar_corr} for the corrected error covariance. In particular, this inconsistency may result in a non-symmetric covariance matrix. To address this, we adopt the more general Joseph’s formula~\cite{Bucy05}, applicable to any linear state-space estimator with white noise, regardless of Gaussianity:
\begin{equation}
	\label{eq:joseph_formula}
	\mathbf{P}_{t \vert t} = (\mathbf{I} - \mathbf{K}_t\mathbf{H}_t)\mathbf{P}_{t \vert t-1}\transpose{(\mathbf{I} - \mathbf{K}_t\mathbf{H}_t)} + \mathbf{K}_t\mathbf{R}_t\mathbf{K}_t.
\end{equation}

Substituting the predicted error covariance from equation~\eqref{eq:covar_pred} into equation~\eqref{eq:joseph_formula}, we express $\mathbf{P}_{t \vert t}$ as the sum of two terms $\mathbf{P}_{t \vert t} = \mathbf{A}_t + \mathbf{B}_t$, where $\mathbf{A}_t$ is the \emph{corrected propagated error covariance}:
\begin{equation}
	\label{eq:propagated_covar}
	\mathbf{A}_t = (\mathbf{I} - \mathbf{K}_t\mathbf{H}_t)\mathbf{F}_t\mathbf{P}_{t-1 \vert t-1}\transpose{\mathbf{F}_t}\transpose{(\mathbf{I} - \mathbf{K}_t\mathbf{H}_t)},
\end{equation}
and $\mathbf{B}_t$ is the \emph{corrected noise covariance}:
\begin{equation}
	\label{eq:corr_noise_covar}
	\mathbf{B}_t = (\mathbf{I} - \mathbf{K}_t\mathbf{H}_t)\mathbf{Q}_t\transpose{(\mathbf{I} - \mathbf{K}_t\mathbf{H}_t)} + \mathbf{K}_t\mathbf{R}_t\transpose{\mathbf{K}_t}.
\end{equation}
The term $\mathbf{A}_t$ is computed in closed form, as it depends only on the gain estimated by the first RNN at time $t$ and the covariance estimated by the second RNN at $t-1$. This recursive dependence on the previously estimated covariance gives rise to the method’s name: Recursive KalmanNet (RKN). In contrast, $\mathbf{B}_t$ is estimated by the second RNN, where we directly learn its Cholesky factor to ensure positive semi-definiteness, \emph{i.e.}, $\mathbf{B}_t = \mathbf{C}_t \transpose{\mathbf{C}_t}$, where $\mathbf{C}_t$ is a lower triangular matrix with real entries.

\subsection{Training and loss function}
The RKN model is trained in a supervised manner using time series datasets consisting of state and measurement pairs, denoted as $(\mathbf{x}_t^{(i)}, \mathbf{z}_t^{(i)})$, where $i \in \{1, \ldots, N\}$ is the batch index, and $t \in \{1, \ldots, T\}$ is the time index. Training is performed with the ADAM optimizer, minimizing a loss function based on the negative Gaussian log-likelihood of the estimation error as in~\cite{UncertaintyKalmanNet}.

Let $\Theta_1$ and $\Theta_2$ denote the parameters of the two RNNs illustrated in Fig.~\ref{fig:architecture}. At each index $i$ and time step $t$, the estimation error depends on $\Theta_1$, while its covariance is influenced by both $\Theta_1$ and $\Theta_2$.
The loss function is defined as:
\begin{alignat}{3}
	\label{eq:logLH_cost}
	\mathcal{L}_t^{(i)}(\Theta_1, \Theta_2) & = &   & \transpose{\mathbf{e}_t^{(i)}(\Theta_1)}
	{\widehat{\mathbf{P}}_{t \vert t}^{(i)}(\Theta_1, \Theta_2)}^{-1}
	\mathbf{e}_t^{(i)}(\Theta_1) \notag \\
	                                        &   & + & \log\det\widehat{\mathbf{P}}_{t \vert t}^{(i)}(\Theta_1, \Theta_2).
\end{alignat}
The overall loss function is computed as the batch and time average of the individual losses $\mathcal{L}_t^{(i)}$, with $\ell_2$ regularization applied to both parameters $\Theta_1$ and $\Theta_2$. These parameters are optimized jointly, rather than in an alternating fashion as in~\cite{CholeskyKalmanNet}.

For each $i$ and $t$, the gradient of the loss function with respect to the gain $\widehat{\mathbf{K}}_t^{(i)}$ and the error covariance $\widehat{\mathbf{P}}_{t \vert t}^{(i)}$ is derived using standard matrix calculus:
\begin{subequations}
	\begin{alignat}{3}
		\label{eq:logLH_gradient_gain}
		\frac{\partial\mathcal{L}_t^{(i)}}{\partial\widehat{\mathbf{P}}_{t \vert t}^{(i)}}
		  & = & {} & {\widehat{\mathbf{P}}_{t \vert t}^{(i)}{}}^{-1} - {\widehat{\mathbf{P}}_{t \vert t}^{(i)}{}}^{-1}\mathbf{e}_t^{(i)}\transpose{\mathbf{e}_t^{(i)}}{\widehat{\mathbf{P}}_{t \vert t}^{(i)}{}}^{-1},                                                                               \\
		\label{eq:logLH_gradient_covar}
		\frac{\partial\mathcal{L}_t^{(i)}}{\partial\widehat{\mathbf{K}}_t^{(i)}}
		  & = & {} & 2\bigg(\frac{\partial\mathcal{L}_t^{(i)}}{\partial\widehat{\mathbf{P}}_{t \vert t}^{(i)}}\cdot \left(\mathbf{I}-\widehat{\mathbf{K}}_t^{(i)}\mathbf{H}_t\right)\mathbf{F}_t\widehat{\mathbf{P}}_{t-1 \vert t-1}^{(i)}\transpose{\mathbf{F}_t}\transpose{\mathbf{H}_t} \nonumber \\
		  &   & -  & {\widehat{\mathbf{P}}_{t \vert t}^{(i)}{}}^{-1}\mathbf{e}_t^{(i)}\transpose{\widehat{\mathbf{y}}_t^{(i)}}\bigg).
	\end{alignat}
\end{subequations}
Therefore, the critical points of $\mathcal{L}_t^{(i)}$ satisfy $\widehat{\mathbf{P}}_{t \vert t}^{(i)} = \mathbf{e}_t^{(i)}\transpose{\mathbf{e}_t^{(i)}}$ and $\mathbf{e}_t^{(i)}\transpose{\widehat{\mathbf{y}}_t^{(i)}} = 0$, yielding orthogonality between the state error $\mathbf{e}_t^{(i)}$ and innovation $\widehat{\mathbf{y}}_t^{(i)}$. The batch averaging in the global loss function guarantees that the learned covariance accurately reflects the true estimation error statistics, while the learned gain decorrelates the estimation error from the innovation. This is a critical condition for optimal filtering, ensuring that all available measurement information is fully utilized to correct the predicted state estimate.

\section{Numerical assessments}
\label{S:Results}
We evaluate RKN performance on a 1D constant-speed linear model, with a position measurement. The state vector, consisting of position and velocity, and the measurement are described by the equations below, where $\dt$ is the time step:
\begin{subequations}
	\begin{alignat}{2}
		\label{eq:transition_model_results}
		\mathbf{x}_{t} & =
		\begin{bmatrix}
			1 & \dt \\
			0 & 1
		\end{bmatrix}
		\mathbf{x}_{t-1} + \mathbf{v}_t, & \quad & \mathbf{x}_{t} \in \mathbb{R}^2, \\
		\label{eq:observation_model_results}
		z_t & =
		\begin{bmatrix}
			1 & 0
		\end{bmatrix}
		\mathbf{x}_{t} + w_t,            & \quad & z_{t} \in \mathbb{R}.
	\end{alignat}
\end{subequations}
The process noise $\mathbf{v}_t$ is zero-mean Gaussian white noise with covariance $\mathbf{Q}_t = \begin{bmatrix} 0 & 0 \\ 0 & {\sigma_v}^2 \end{bmatrix}$, modeling white noise on acceleration. This formulation leads to a velocity random walk characterized by zero-mean Gaussian increments with variance ${\sigma_v}^2$. The measurement noise $w_t$ follows a heavy-tailed bimodal-Gaussian distribution:
\[w_t = Z_t X_t + (1 - Z_t) Y_t,\]
where $X_t$, $Y_t$, and $Z_t$ are independent white noise processes, with $X_t$ and $Y_t$ being Gaussian with variances ${\sigma_1}^2$ and ${\sigma_2}^2$, respectively, and $Z_t$ is Bernoulli with parameter $p$. Then, $w_t$ is distributed as $p \mathcal{N}(0, {\sigma_1}^2) + (1 - p) \mathcal{N}(0, {\sigma_2}^2)$. It thus has variance $\mathbf{R}_t = Z_t {\sigma_1}^2 + (1 - Z_t) {\sigma_2}^2$, and an expected variance of ${\sigma_w}^2 = p{\sigma_1}^2 + (1 - p){\sigma_2}^2$.

Finally, we set $\dt = 1$ without units and define the noise heterogeneity level as $\nu = \dfrac{{\sigma_w}^2}{{\sigma_v}^2}$.

We compare RKN with two versions of the KF: the optimal KF (o-KF) and the sub-optimal KF (so-KF). The o-KF provides the optimal gain and state estimate since it has access to the true noise covariance, $\mathbf{R}_t$, at each time step, making it the reference method. In real-world scenarios, the noise covariance is usually unknown, so an approximate Gaussian model is often used. The so-KF approximates the noise covariance with a single Gaussian of variance ${\sigma_w}^2$. Furthermore, we compare RKN with Cholesky KalmanNet (CKN)~\cite{CholeskyKalmanNet}, a recent KalmanNet-derived method that outperforms~\cite{KalmanNet,SplitKalmanNet}.

We train the RKN model using the squared features from Section~\ref{ss:features} on synthetically generated time series based on equations~\eqref{eq:transition_model_results} and~\eqref{eq:observation_model_results}. Each series has $T=150$ samples, with $1000$ time series for training, $100$ for validation, and $1000$ for testing. Initial conditions follow a normal distribution with mean $\widehat{\mathbf{x}}_{0\vert0}$ and covariance $\widehat{\mathbf{P}}_{0\vert0}$. The Bernoulli sampling process varies across series.

For our experiments, we set the initial error covariance as $\widehat{\mathbf{P}}_{0 \vert 0} = \begin{bmatrix} 1 & 0 \\ 0 & 0.01 \end{bmatrix}$ and the initial mean state as $\widehat{\mathbf{x}}_{0 \vert 0} = \begin{bmatrix} 0 \\ 1 \end{bmatrix}$. We further set ${\sigma_v}^2 = 10^{-4}$, $p = 0.6$, and ${\sigma_1}^2 = 1.5625{\sigma_w}^2$. Note that the observations discussed below also hold under different parameter settings.

\subsection{Performance at varying noise heterogeneity level}

\begin{table}[tbp!]
    \centering
    \resizebox{\columnwidth}{!}{%
    \begin{tabular}{|c|c|c|c|c|c|c|c|c|c|c|}
        \hline
        \textbf{$\nu$} & \multicolumn{2}{c|}{20} & \multicolumn{2}{c|}{30} & \multicolumn{2}{c|}{40} & \multicolumn{2}{c|}{50} & \multicolumn{2}{c|}{60} \\
        \specialrule{1.2pt}{0pt}{0pt}
        \textbf{o-KF} & $-28$ & $2.0$ & $-21$ & $2.0$ & $-14$ & $2.0$ & $-6.9$ & $2.0$ & $0.1$ & $2.0$\\
        \hline
        \textbf{so-KF} & $-26$ & $2.0$ & $-18$ & $2.0$ & $-11$ & $2.0$ & $-3.9$ & $2.0$ & $3.4$ & $2.0$\\
        \hline
        \textbf{CKN} & $-20$ & $4.0$ & $-17$ & $27$ & $-11$ & $3.2$ & $-4.8$ & $6.2$ & $2.4$ & $21$\\
        \hline
        \textbf{RKN} & $-26$ & $2.0$ & $-19$ & $2.0$ & $-12$ & $2.0$ & $-5.1$ & $1.9$ & $2.3$ & $2.2$\\
        \hline
    \end{tabular}%
    }
    \caption{MSE (in dB, left) and MSMD (right) for varying noise heterogeneity levels, $\nu$ [dB].}
    \label{T:MSE_MSMD}
\end{table}

\begin{figure*}[ht]
	\centering
	\subfloat[]{
		\includegraphics[width=0.43\textwidth]{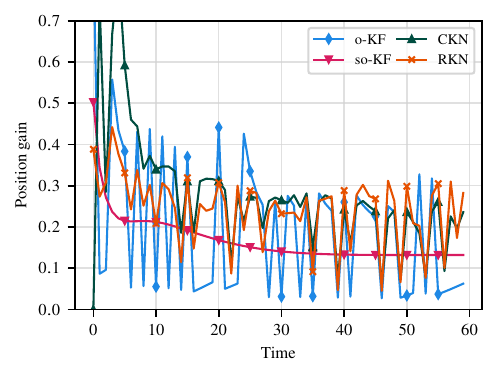}
		\label{sfig:gain}
	}
	\hspace{0.1cm}
	\subfloat[]{%
		\includegraphics[width=0.43\textwidth]{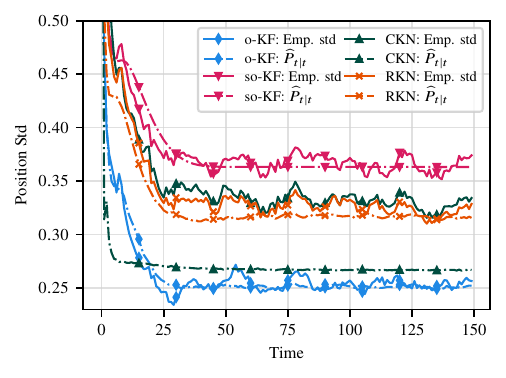}
		\label{sfig:P}
	}
	\caption{Comparison of position estimates. (a) Gain from a single time series (zoom on the first 60 samples). (b) Root mean square values over 1000 test time series of the estimated (solid lines) and empirical (dash-dot lines) standard deviations.}
	\label{fig:temporal}
\end{figure*}
In the following experiment, the noise heterogeneity level is varied between $20$~dB and $60$~dB and the estimators are assessed using two metrics: Mean Squared Error (MSE) for state estimation accuracy and Mean Squared Mahalanobis Distance (MSMD) for the consistency of error covariance estimation with state error. These metrics are defined as follows:
\begin{align*}
	\text{MSE}  & = \frac{1}{T} \sum_{t=1}^{T} \frac{1}{N} \sum_{i=1}^{N} \transpose{\mathbf{e}_t^{(i)}} \mathbf{e}_t^{(i)},                               \\
	\text{MSMD} & = \frac{1}{T} \sum_{t=1}^{T} \frac{1}{N} \sum_{i=1}^{N} \transpose{\mathbf{e}_t^{(i)}} {\mathbf{P}_{t|t}^{(i)}}^{-1} \mathbf{e}_t^{(i)}.
\end{align*}
If $\mathbf{e}_t^{(i)}$ is Gaussian, then $\transpose{\mathbf{e}_t^{(i)}} \mathbf{P}_{t|t}^{(i)^{-1}} \mathbf{e}_t^{(i)}$ follows a chi-squared distribution with $m$ degrees of freedom, where $m$ is the state dimension. According to the Central Limit Theorem, the batch average of this quantity converges to a Gaussian distribution with mean $m$ and variance $2m/N$. The Gaussian nature of state errors is an inherent property of the optimal KF, but this characteristic is also highly desirable for other estimators, as it ensures that all available information is incorporated into the estimate. However, due to temporal dependencies in the state error, deriving the distribution of the MSMD becomes considerably more complex, although its mean remains $m$.

Table~\ref{T:MSE_MSMD} presents the MSE and MSMD obtained for the compared methods. The results indicate that RKN outperforms both so-KF and CKN in terms of MSE, achieving performance closest to the reference values produced by o-KF.
The MSMD values for both KF and RKN closely match the theoretical mean of $m=2$. In contrast, CKN yields significantly higher MSMD values, reflecting inconsistency in its covariance representation. This can be traced to CKN’s loss function, which uses a hyperparameter to balance the MSE and covariance terms~\cite[Equation~(7)]{CholeskyKalmanNet}. In the authors’ implementation, the covariance term is weighted at just 0.05, with 0.95 assigned to the MSE, favoring low MSE at the expense of covariance accuracy. RKN, by contrast, uses a tuning-free loss, inherently balancing both aspects without manual adjustment.

\subsection{Temporal analysis of gain and error covariance estimations}
To further analyze into RKN’s performance, we analyze the temporal evolution of the gain and covariance estimates under a fixed $\nu = 40$ dB, yielding ${\sigma_w}^2 = 1$.
Fig.~\ref{sfig:gain} shows that the optimal position gain from o-KF exhibits a noisy, time-varying behavior driven by the Bernoulli process switching Gaussian modes. Both RKN and CKN partially track this dynamic, while so-KF—as expected—produces a smoother, less responsive gain profile.
In Fig.~\ref{sfig:P} we assess position estimate accuracy via the standard deviations shown by the dash-dot lines, and evaluate consistency between the estimated and empirical standard deviations of the position errors by comparing the dash-dot and solid lines. RKN and CKN yield covariances between the optimal o-KF and the suboptimal so-KF. However, only RKN’s covariance estimates consistently reflect the actual error spread, while CKN underestimates it. Similar trends are observed for velocity estimates.

These findings demonstrate that RKN reliably estimates the states, gain, and error covariance, even under challenging non-Gaussian measurement noise. This paper does not address the model's generalization capabilities, which are explored separately in~\cite{Falcon25}, where performance is evaluated under a mismatch between training and test noise variance ranges. To enhance the generalization capabilities of the learned covariance, future work will focus on analyzing the learned Cholesky factor $\mathbf{C}_t$ 
to ensure its consistency with the learned gain $\mathbf{K}_t$.

\section{Conclusion}
\label{S:Conclusion}
This paper introduces RKN, a deep-learning augmented Kalman filter that leverages two recurrent neural networks to estimate gain and error covariance. The latter is recursively computed using a formulation derived from Joseph’s equation. In scenarios with bimodal Gaussian noise, RKN outperformes conventional Kalman filters and a state-of-the-art deep learning approach, delivering accurate state and covariance estimates that closely reflect actual errors. RKN shows strong potential in cases where classical Kalman filtering falls short, such as in non-linear systems or with incomplete state-space models.

\renewcommand{\refname}{\footnotesize References}
\footnotesize
\bibliography{biblio.bib}
\bibliographystyle{IEEEtran}

\end{document}